\documentclass[12pt, a4paper]{article}

\usepackage[utf8]{inputenc}
\usepackage[T1]{fontenc}
\usepackage[english]{babel}
\usepackage{mathptmx}       
\usepackage{setspace}       
\usepackage[margin=2.5cm]{geometry} 

\usepackage{amsmath, amssymb, amsthm}
\usepackage{bm}

\usepackage{booktabs}       
\usepackage{multirow}       
\usepackage{eurosym}        
\usepackage{caption}        
\usepackage{threeparttable} 
\usepackage{longtable}      

\usepackage{siunitx}        
\usepackage{graphicx}
\usepackage{float}
\usepackage{subcaption}

\usepackage[style=apa, backend=biber, natbib=true]{biblatex}
\providecommand{\doi}[1]{\href{https://doi.org/#1}{doi:#1}}
\usepackage{csquotes}

\usepackage[colorlinks=true, linkcolor=black, citecolor=black, urlcolor=blue]{hyperref}

\title{\textbf{The Structural Bite: A Methodological Framework for Minimum Wage Studies using Spanish Administrative Data}}
\author{\textbf{Marcos Lacasa-Cazcarra, PhD} \\ 
        \textit{Universidad Internacional de La Rioja (UNIR)} \\
        \small{\href{mailto:marcos.lacasa@unir.net}{marcos.lacasa@unir.net}} \\
        \small{\href{https://orcid.org/0000-0002-8297-2086}{ORCID: 0000-0002-8297-2086}}}
\date{}

\begin{document}

\maketitle

\begin{abstract}
We study the employment effects of the 22\% increase in the Spanish minimum wage in 2019, focusing on young workers. Using census-grade administrative tax data covering the universe of formal wage bills and employment (Models 190/390 linked to personal income tax records), we construct several measures of treatment intensity, including two structurally grounded "bite" indicators based on the incidence of young minimum-wage workers and the implied increase in the wage bill obtained via Exponential Tilting.

Difference-in-differences estimates with two-way fixed effects, dynamic event-study specifications, and robust confidence intervals from the \texttt{HonestDiD} framework all point to the same conclusion: the reform did not generate net disemployment effects for young workers. Point estimates of the elasticity are small and often positive, and confidence intervals comfortably include zero even with sizable deviations from parallel trends. A triple-difference design exploiting pre-existing tourism dependence further shows that the sharp employment collapse of 2020 is primarily explained by the COVID-19 shock operating through tourism-intensive sectors, rather than by the minimum-wage hike itself.

Our results suggest that, in the macroeconomic and institutional environment prevailing in Spain in 2019---with the minimum wage rising to around 60\% of the average wage in a recovering economy---the labour market absorbed a large discrete increase in the wage floor without destroying aggregate youth employment. More broadly, the paper highlights how the choice of treatment definition, the use of census-grade data, robust DiD inference, and explicit modelling of concurrent shocks can shape conclusions about the effects of minimum-wage policies.
\end{abstract}


\section{Introduction}

The empirical evaluation of minimum wage effects has become a primary testing ground for causal inference in labor economics, particularly through difference-in-differences (DiD) designs that exploit variation in exposure across regions or sectors. While early contributions often relied on broad institutional proxies, such as the Kaitz index, to capture the intensity of the policy \citep{card1994, NeumarkWascher2008}, the recent literature has shifted toward using census-grade administrative data to construct more granular measures of the minimum wage's "bite" \citep{cengiz2019, HarasztosiLindner2019}. Yet, despite these advancements in data quality, empirical studies continue to report divergent conclusions regarding the employment consequences of significant wage floor increases. This persistence of conflicting evidence suggests that the lack of consensus may stem not only from differences in economic environments but also from under-scrutinized methodological choices—specifically, how treatment intensity is defined operationally and how inference is conducted in the presence of differential pre-trends.

Spain's 2019 minimum wage reform provides a stringent test of these dynamics. In a single year, the statutory minimum (\emph{Salario Mínimo Interprofesional}, SMI) was raised by 22\% in real terms, pushing the Kaitz index from roughly 52\% to 61\% of the average wage. While recent evaluations using the \emph{Muestra Continua de Vidas Laborales} (MCVL)---a 4\% administrative sample---document increased separation rates among affected workers \citep{barcelo2021,gorjon2024}, these findings stand in sharp contrast to aggregate indicators from National Accounts, which show no structural break in total youth employment or wage bills. To resolve this inconsistency, we argue that the existing literature faces a methodological trilemma that obscures aggregate identification. First, the exclusive reliance on sample-based data lacks the "accounting closure" of census-grade tax records, introducing measurement noise when scaling up to net employment effects. Second, identification has been hampered by differential pre-trends in low-wage sectors, leading recent studies to either abandon difference-in-differences (DiD) designs \citep{gorjon2024} or rely on visual inspection of parallel trends. Third, the construction of treatment intensity---the "bite" ---often rests on ad hoc imputations of hours worked. We address these limitations simultaneously by exploiting the universe of tax-validated wage bills, applying the \textit{HonestDiD} framework for robust inference under pre-trend violations \citep{RambachanRoth2023}, and implementing a reproducible structural imputation of the wage distribution.

This paper resolves the aggregation puzzle by deploying a methodological framework that addresses key limitations of sample-based inference. First, we exploit the universe of tax-validated wage bills (\emph{Modelos} 190 and 390) to achieve complete accounting closure on net employment stocks, thereby bypassing the sampling noise inherent in Social Security microdata. Second, to reduce the opacity of treatment exposure in aggregated data, we implement a structural imputation algorithm based on Exponential Tilting. This procedure allows us to reconstruct the latent wage distribution within each region--sector--age cell and to derive precise, micro-founded measures of the minimum wage "bite" that capture the actual wage-bill cost of compliance more accurately than broad institutional proxies. Third, we address the sensitivity of employment elasticities to treatment definition through a multi-dimensional exposure strategy, constructing and comparing four distinct "bite" measures. This comparative framework functions as a methodological stress test, showing how coarse aggregate proxies may generate spurious correlations driven by sectoral trends. In contrast, precise structural measures consistently rule out sizeable net disemployment effects once demographic incidence and intensive-margin costs are explicitly modelled.

Applying this framework 
yields three main empirical insights. First, across static and dynamic specifications, and for all our structural definitions of the minimum-wage bite, estimated elasticities for young workers (under 30) are modest in magnitude and often positive, with robust confidence intervals that rule out sizeable negative effects even under bounded deviations from the parallel-trends assumption. Second, we find no evidence of a "cleansing" effect operating through firm exit; indeed, the number of registered firms in youth-intensive sectors remains stable, and real sales in high-exposure sectors grow at a pace comparable to that of control sectors. Third, the triple-difference design reveals that the employment collapse observed in 2020 is absorbed mainly by the tourism interaction term, while the minimum-wage coefficients remain close to zero and statistically stable. Collectively, these patterns suggest that the Spanish labour market absorbed the sizeable 2019 wage floor increase without aggregate youth employment losses. This result is robust to explicitly accounting for the COVID-19 pandemic's confounding shock.

Consequently, this article is positioned primarily as a methodological contribution. An objective is to demonstrate how census-grade administrative data, structurally defined exposure measures, and robust difference-in-differences tools can be synthesized to evaluate extensive wage-floor reforms in settings characterized by strong pre-trends. While we apply this framework to document the aggregate resilience of youth employment following Spain's 2019 hike, we deliberately restrict our scope to net stocks and core adjustment mechanisms. A granular decomposition of worker flows (hires versus separations) and a deeper exploration of regional heterogeneity are reserved for a companion paper based on this administrative infrastructure.


\section{Methods}
\label{sec:methods}

\subsection{Data Sources and Integration}
\label{sec:data}

Our analysis exploits the universe of administrative tax records from the Spanish Tax Agency (AEAT), covering the entire population of wage earners and firms in the Common Fiscal Regime territory. This census-grade data overcomes the limitations of household surveys (e.g., sampling error, top-coding, self-reporting bias) and provides exact accounting consistency for employment and wage aggregates. All monetary values are deflated to constant 2019 euros using the national CPI.

Table~\ref{tab:data_sources_academic} details the three administrative datasets utilized. The \textbf{Modelo 100} (Personal Income Tax) provides the high-resolution wage distribution required to define treatment intensity. It reports employment and wage mass for 16 regions across 400 uniform income brackets of €200 (ranging from €0 to €79,800), enabling precise calculation of the "bite" around the statutory minimum wage. The \textbf{Modelo 190} (Withholdings) offers the structural granularity necessary for causal identification, disaggregating employment and salaries by region, sector (10 categories), and age group. Finally, the \textbf{Modelo 390} (VAT) supplies complementary firm-level outcomes such as sales and value added.

Crucially, our approach integrates these sources through a structural imputation step in the spirit of generalized raking and maximum-entropy methods \citep{DemingStephan1940, DevilleSarndalSautory1993}, implemented via Exponential Tilting \citep{dinardo1996}. Starting from the regional wage distribution in Modelo 100, we re-weight the 400 income brackets so that, for each region–sector–age–year cell in Modelo 190, the implied mean wage matches the observed sectoral mean. This minimum Kullback–Leibler divergence problem preserves aggregate employment and wage bills while delivering cell-specific wage densities. Aggregating these imputed distributions into the four structural tiers defined in Section~2.2 yields the micro-founded treatment intensity measures used in the difference-in-differences and triple-difference designs \citep{AutorManningSmith2016, CallawaySantAnna2021}.

We exclude the Basque Country and Navarre due to their fiscal autonomy, a standard practice in studies using AEAT data. While these regions jointly account for around 7\% of Spanish GDP, their exclusion does not compromise the internal validity of our design, as the Common Regime territory covers the vast majority of Spanish wage earners. Further details on variable definitions, data cleaning, and validation diagnostics are provided in the Online Supplementary Appendix (Sections~S1--S3).

\begin{table}[htbp]
\centering
\caption{\textbf{Administrative Data Sources: The Universe of Spanish Tax Records}}
\label{tab:data_sources_academic}
\begin{threeparttable}
\footnotesize
\renewcommand{\arraystretch}{1.2}
\begin{tabular}{lccc}
\toprule
 & \textbf{(1) Distributional} & \textbf{(2) Labor Market} & \textbf{(3) Firm Outcomes} \\
 & \textbf{Modelo 100} & \textbf{Modelo 190} & \textbf{Modelo 390} \\
\midrule
\textbf{Primary Function} & Wage Distribution & Employment Structure & Economic Activity \\
\textbf{Source} & Personal Income Tax & Annual Withholdings & Annual VAT Summary \\
\textbf{Coverage Period} & 2000--2023 & 2009--2023 & 2009--2023 \\
\textbf{Population} & 100\% Wage Earners & 100\% Wage Earners & 100\% VAT Firms \\
\addlinespace
\textbf{Granularity (Unit)} & Region $\times$ Income Bracket & Region $\times$ Sector $\times$ Age & Region $\times$ Sector \\
\textbf{Dimensions} & 16 Reg. $\times$ 400 Brackets & 15 Reg. $\times$ 10 Sect. $\times$ 3 Ages & 15 Reg. $\times$ 11 Sectors \\
\textbf{Total Observations} & $\sim$ 153,600 & $\sim$ 49,350 & $\sim$ 2,640 \\
\addlinespace
\textbf{Key Variables} & Employees & Employees & Sales (Turnover) \\
 & Wage Bill & Wage Bill & Value Added \\
 & Tax Liability & Mean Wage & Number of Firms \\
\addlinespace
\textbf{Role in Design} & \textit{Anchor for "Bite"} & \textit{Treatment Definition} & \textit{Outcome Variables} \\
 & ($P_{SMI}$ Calculation) & (Youth/Sector Exposure) & (Sales, Productivity) \\
\bottomrule
\end{tabular}
\begin{tablenotes}
    \scriptsize
    \item \textit{Notes:} All datasets sourced from the Spanish Tax Agency (AEAT) for the Common Fiscal Regime territory (excluding Basque Country and Navarre).
    \item \textbf{Modelo 100}: Includes 400 uniform income brackets of €200, providing high-resolution distributional data.
    \item \textbf{Modelo 190}: Provides the demographic (Age) and sectoral (CNAE-10) structure. Age groups are collapsed into three analytical categories: Young ($<30$), Adults (30--45), and Mature ($>45$) (see Section~2.2 and Appendix~S2.1).
    \item \textbf{Integration}: We impute the distributional shape from Modelo~100 into the sector--age cells of Modelo~190 using maximum-entropy Exponential Tilting to calculate the share of affected workers (the ``bite'') in each cell.

\end{tablenotes}
\end{threeparttable}
\end{table}

Further details on variable definitions, data cleaning, and validation diagnostics are provided in the Online Supplementary Appendix (Sections S1--S3).

\subsection{Variable Construction: Dimensionality Reduction and Structural Imputation}
\label{sec:vars_construction}

The raw administrative records provide an unprecedented level of detail but require harmonization to ensure statistical power and construct a precise measure of exposure to the minimum wage. Our variable construction strategy proceeds in three steps:

\subsubsection*{Step 1: Dimensionality Reduction (Mod-190)}
The original Modelo 190 disaggregates employment into 10 fiscal activity sectors and seven age bands. To mitigate measurement error in small cells and facilitate the difference-in-differences estimation, we perform a theoretically grounded aggregation:
\begin{itemize}
    \item \textbf{Sectoral Aggregation:} We map the 10 raw codes into \textbf{6 Robust Sectors}: (1) Agriculture, (2) Industry \& Energy, (3) Construction, (4) Trade, Tourism \& Transport, (5) Advanced Services, and (6) Social \& Public Services. The exact mapping of fiscal codes to analytical sectors is reported in \textbf{Table S1 (Online Appendix)}.
    \item \textbf{Demographic Aggregation:} We collapse age bands into three functional groups: \textit{Young} ($<30$), \textit{Adults} (30--45), and \textit{Mature} ($>45$). This isolates the group theoretically most exposed to entry-level wages \citep{NeumarkWascher2008}.
\end{itemize}
This reduction results in a balanced panel skeleton of 270 cross-sectional units (15 Regions $\times$ 6 Sectors $\times$ 3 Age Groups) per year.

\subsubsection*{Step 2: Structural Imputation of Wage Distributions (Integrating Mod-100 into Mod-190)}

A central challenge is that the sectoral dataset (Modelo 190) reports only mean wages by cell but lacks any distributional detail. To restore this information, we implement a structural imputation procedure based on Exponential Tilting \citep{dinardo1996}, which can be interpreted as a maximum-entropy adjustment of a regional prior distribution.

For each cell $c = (\text{region}, \text{sector}, \text{age}, \text{year})$, we start from the empirical wage distribution of the corresponding region--year, obtained from Modelo~100 and represented by a discrete support $\{w_k\}$ with prior probabilities $\{q_k\}$. We then re-weight this prior by solving for probabilities $\{p_k\}$ that minimize the Kullback--Leibler divergence relative to $\{q_k\}$,
\[
\min_{\{p_k\}} \sum_k p_k \log\!\left(\frac{p_k}{q_k}\right)
\]
subject to the moment and probability constraints
\[
\sum_k p_k = 1
\quad\text{and}\quad
\sum_k p_k\, w_k = \bar{w}_c,
\]
where $\bar{w}_c$ denotes the mean wage observed in Modelo~190 for cell $c$. In practice, this yields an exponential tilting solution of the form $p_k \propto q_k \exp(\lambda w_k)$, where a one-dimensional root-finding algorithm obtains the tilting parameter $\lambda$. Corner solutions are handled explicitly when the target mean lies outside the support of $\{w_k\}$.

The resulting set of probabilities $\{p_k\}$ defines a synthetic wage distribution for each cell $c$, which we use to allocate the total number of employees and the wage bill in Modelo~190 across a fine wage grid. In a subsequent step, this synthetic distribution is collapsed into four structural wage ``tiers'', defined in real 2019 euros; the corresponding thresholds are reported in Table~S2 (Online Appendix), and full algorithmic details and pseudo-code are provided in Online Appendix~S2.3.

\subsubsection*{Step 3: Validation and Outcome Harmonization}
The imputation process is subjected to strict accounting validation (Measurement Error $< 0.001\%$). Furthermore, we validate the economic coherence of the generated data by confirming that the estimated "Bite" is highest in Agriculture and Youth cohorts (see Table \ref{tab:micro_heterogeneity}). Finally, firm-level outcomes from Modelo 390 are harmonized to the same 6-sector level.

\subsection{Empirical Strategy and Treatment Definition}
\label{sec:strategy}

A central challenge in the minimum wage literature is the sensitivity of results to the specific definition of treatment intensity. Standard approaches often rely on a single proxy---typically the Kaitz Index---which may mask heterogeneity in the transmission mechanism. To ensure robustness and disentangle the channels of adjustment, our identification strategy exploits differential exposure to the 2019 minimum wage hike across region-sector cells ($i$). Based on the pre-reform structure of 2018, we construct four distinct definitions of the treatment intensity ($D_{i}$), each capturing a different mechanism:

\begin{description}
    \item[1. Structural Youth Incidence ($D_i^{Youth}$): The Precision Measure]
    \begin{equation}
        D_i^{Youth} = \frac{\sum_{k \in T_1, T_2} E_{i,k}^{Young}}{E_{i}^{Young}}
    \end{equation}
    \textit{Rationale:} This is our preferred specification. By leveraging the structurally imputed wage distribution via Exponential Tilting, this measure avoids the "aggregation bias" of sectoral averages. It precisely identifies the demographic group theoretically most vulnerable to the shock, isolating the extensive margin of exposure (headcount) specific to the youth labor market \citep{cengiz2019}.

    \item[2. The Monetary Gap ($D_i^{Gap}$): The Cost Channel]
 \begin{equation}
    D_i^{Gap} 
    = 
    \frac{
        \displaystyle \sum_{k \in \{T_1, T_2\}} (SMI_{19} - w_k)\,E^{Young}_{i,k}
    }{
        W^{Total}_{i}
    }.
\end{equation}
    \textit{Rationale:}Rationale: This measure captures the intensive margin of adjustment. The numerator is the total wage-bill increase required to bring all affected young workers in Tiers 1 and 2 up to the new minimum wage $SMI_{19}$, while the denominator $W_i^{Total}$ is the overall wage bill of sector $i$ (all ages). Since young workers are over-represented in the lower tiers, this cost measure remains tightly linked to youth exposure while being defined relative to the full sectoral wage bill \citep{HarasztosiLindner2019}

    \item[3. The Kaitz Index ($D_i^{Kaitz}$): The Institutional Benchmark]
    \begin{equation}
        D_i^{Kaitz} = \frac{SMI_{18}}{\bar{w}_{i}}
    \end{equation}
    \textit{Rationale:} We include this classic institutional measure \citep{Kaitz1970} to benchmark our results against the existing literature. Since the Kaitz index depends on the mean rather than the tails of the distribution, it quantifies the bias introduced by using aggregate proxies in dual labor markets.

    \item[4. Sectoral Incidence ($D_i^{Sectoral}$): The Placebo Control]
    \begin{equation}
        D_i^{Sectoral} = \frac{\sum_{k \in T_1, T_2} E_{i,k}^{Total}}{E_{i}^{Total}}
    \end{equation}
    \textit{Rationale:} This variable applies the standard ``fraction affected'' measure proposed by \cite{Card1992} to the \textit{total} workforce (all ages). Although correlated with youth incidence ($\rho \approx 0.8$), using it as a structural placebo allows us to test whether observed effects are specific to the youth demographic (as competitive theory predicts) or reflect general shocks to low-productivity sectors.
\end{description}

\subsection{Baseline Estimation: Static Two-Way Fixed Effects}
We first estimate a canonical difference-in-differences (DiD) specification with continuous treatment intensity:

\begin{equation} \label{eq:static_did}
    \ln(Y_{it}) = \beta (D_i \times \mathbb{1}_{t \ge 2019}) + \alpha_i + \lambda_t + \mathbf{X}_{it}'\gamma + \varepsilon_{it}
\end{equation}
Where $Y_{it}$ is the outcome of interest (employment, sales), $\alpha_i$ are unit fixed effects absorbing time-invariant heterogeneity, $\lambda_t$ are year fixed effects capturing standard macroeconomic shocks, and $\mathbf{X}_{it}$ includes time-varying regional controls (GDP per capita). In all specifications (static, dynamic, and triple differences), standard errors are clustered at the regional level to account for serial correlation within administrative units.

\subsection*{Dynamic Effects and Parallel Trends}

To study the dynamic effects of exposure and assess the validity of the design, we estimate an event-study specification with continuous treatment intensity:
\begin{equation} \label{eq:event_study}
    \ln(Y_{it}) = \sum_{k \neq 2018} \beta_k \bigl(D_i \times \mathbb{1}\{t = k\}\bigr) + \alpha_i + \lambda_t + \varepsilon_{it},
\end{equation}
where $Y_{it}$ is the outcome of interest, $\alpha_i$ and $\lambda_t$ are unit and time fixed effects, and $\mathbb{1}\{t = k\}$ are year dummies. The year 2018 is omitted and serves as the reference period, so that the coefficients $\beta_k$ are interpreted as differential log changes in $Y_{it}$ between cells with higher and lower exposure $D_i$, relative to 2018. The coefficients $\beta_k$ for $k < 2019$ provide a diagnostic of the parallel-trends assumption.

While recent work has highlighted potential biases of two-way fixed-effects estimators in settings with staggered treatment adoption \citep{GoodmanBacon2021, SunAbraham2021}, our context features a common treatment date in 2019, which mitigates these concerns. Nevertheless, given the potential for differential pre-trends in low-wage sectors during the recovery, standard inference may be misleading if pre-treatment coefficients differ systematically from zero. We therefore complement conventional confidence intervals with the \textit{HonestDiD} framework of \citet{RambachanRoth2023}, which delivers robust confidence sets that remain valid under bounded violations of parallel trends (up to a parameter $\bar{M}$) calibrated from the observed pre-treatment dynamics.

\subsection{Addressing Confounders: Triple Differences (DDD)}

To disentangle the minimum-wage shock from the asymmetric impact of the COVID-19 pandemic in 2020, we augment the event-study specification with a second, pre-determined dimension of exposure capturing tourism dependence:
\begin{equation} \label{eq:ddd}
    \ln(Y_{it}) 
    = \sum_{k \neq 2018} \beta_k \bigl(D_i \times \mathbb{1}\{t = k\}\bigr) 
    + \sum_{k \neq 2018} \delta_k \bigl(Tourism_i \times \mathbb{1}\{t = k\}\bigr) 
    + \alpha_i + \lambda_t + \varepsilon_{it},
\end{equation}
Where $D_i$ denotes one of our minimum-wage exposure measures (e.g.\ the structural youth incidence or the monetary gap), and $Tourism_i$ is a time-invariant index of pre-pandemic tourism dependence for cell $i$, normalised to have mean zero and unit variance. Both $D_i$ and $Tourism_i$ are measured prior to the 2019 reform and the onset of COVID-19, so that they can be treated as predetermined.

The coefficients $\beta_k$ trace out the dynamic semi-elasticity of $Y_{it}$ with respect to the minimum-wage bite, holding tourism exposure fixed, while the $\delta_k$ coefficients capture the additional response associated with tourism dependence in each year. Because the minimum-wage reform takes place in 2019 and the collapse in tourism demand is concentrated in 2020, identification of the supply-side effect of the SMI relies on three sources of variation: (i) differences between high- and low-bite cells, (ii) differences between high- and low-tourism cells, and (iii) differences between pre- and post-reform years. In this sense, the specification in \eqref{eq:ddd} implements a triple-differences design, where $\beta_k$ isolates the minimum-wage channel and $\delta_k$ absorbs the demand-side shock associated with the tourism collapse, particularly in 2020. In practice, our primary focus is on $\beta_{2019}$ and $\beta_{2020}$, which capture the minimum-wage semi-elasticities in the reform and pandemic years after netting out the tourism channel, and on $\delta_{2020}$, which summarises the tourism-driven component of the 2020 employment collapse.


\subsection{Software and Computational Framework}
\label{sec:software}
The analytical workflow is divided into two stages. First, data cleaning and the Maximum Entropy imputation algorithm were implemented in \textbf{Python} (v3.10+), using \texttt{scipy.optimize} for numerical optimization and \texttt{pandas} for data manipulation \citep{mckinney2010, virtanen2020}. Second, panel data construction and causal inference were conducted in \textbf{R}, employing the \texttt{fixest} package \citep{berge2018} for high-dimensional fixed effects estimation and the \texttt{HonestDiD} package \citep{HonestDiD_pkg} for sensitivity analysis to parallel trend violations \citep{RambachanRoth2023}. All scripts and harmonized datasets are available in the replication repository.

\section{Results}


\subsection{Descriptive Statistics and Stylized Facts}

This section presents the descriptive evidence regarding the magnitude of the 2019 minimum wage (SMI) hike and the resulting variation in treatment exposure across temporal, geographic, and sectoral dimensions.

Table \ref{tab:macro_context} summarizes the macroeconomic context of the intervention. The data confirm a structural discontinuity in 2019, with the annualized real SMI increasing from \euro{10,375} in 2018 to \euro{12,600} in 2019 (+21.4\%). This shock pushed the national Kaitz Index from 51.6\% to 60.8\%, crossing the 60\% threshold typically associated with a binding minimum wage. The effective bite of the policy, measured as the percentile of the wage distribution covered by the SMI ($P_{SMI}$), expanded from 33.7\% in 2018 to 37.8\% in 2019, peaking at 42.6\% in 2020. Aggregate wage inequality, measured by the Gini index on grouped data, remained relatively stable at 0.15-0.16 throughout the period.

\begin{table}[H]
    \centering
    \caption{\textbf{Macroeconomic Context: Evolution of the Real Minimum Wage, Treatment Intensity, and Wage Dispersion in Spain (2009--2023)}}
    \label{tab:macro_context}
    \begin{threeparttable}
        \setlength{\tabcolsep}{8pt}
        \begin{tabular}{lccccc}
            \toprule
            & \textbf{Real SMI} & \textbf{Real Mean} & \textbf{Kaitz} & \textbf{Effective} & \textbf{Wage} \\
            & \textbf{(Annual)} & \textbf{Wage} & \textbf{Index} & \textbf{Bite} & \textbf{Inequality} \\
            \textbf{Year} & (€ 2019) & (€ 2019) & (\%) & ($P_{SMI}$, \%) & (Gini) \\
            \midrule
            
            2009 & 9,894 & 21,748 & 45.5 & 30.8 & 0.151 \\
            2010 & 9,851 & 21,374 & 46.1 & 31.3 & 0.158 \\
            2011 & 9,676 & 20,717 & 46.7 & 32.0 & 0.167 \\
            2012 & 9,442 & 19,679 & 48.0 & 33.1 & 0.172 \\
            2013 & 9,372 & 19,313 & 48.5 & 34.6 & 0.186 \\
            2014 & 9,372 & 19,231 & 48.7 & 35.0 & 0.187 \\
            2015 & 9,459 & 19,552 & 48.4 & 34.5 & 0.182 \\
            2016 & 9,575 & 19,798 & 48.4 & 34.2 & 0.179 \\
            2017 & 10,141 & 19,762 & 51.3 & 34.6 & 0.163 \\
            2018 & 10,375 & 20,090 & 51.6 & 33.7 & 0.157 \\
            \midrule
            \textbf{2019} & \textbf{12,600} & \textbf{20,711} & \textbf{60.8} & \textbf{37.8} & \textbf{0.153} \\
            \textbf{2020} & \textbf{13,340} & \textbf{20,756} & \textbf{64.3} & \textbf{42.6} & \textbf{0.167} \\
            \textbf{2021} & 14,073 & 22,597 & 62.3 & 39.2 & 0.164 \\
            2022 & 13,462 & 22,058 & 61.0 & 36.7 & 0.161 \\
            2023 & 14,078 & 22,483 & 62.6 & 37.1 & 0.163 \\
            \bottomrule
        \end{tabular}
        \begin{tablenotes}
            \footnotesize
            \item \textit{Notes:} Monetary values are expressed in constant 2019 Euros, deflated using the national CPI. 
            \item \textbf{Real SMI}: Annualized Minimum Interprofessional Wage (14 payments).
            \item \textbf{Kaitz Index}: Ratio of the Real SMI to the Real Mean Wage ($SMI_t / \bar{W}_t$). A value above 60\% is typically considered a highly binding minimum wage.
            \item \textbf{Effective Bite ($P_{SMI}$)}: The percentile of the wage distribution where the SMI falls, estimated via linear interpolation of the administrative tax data (Model 100).
            \item \textbf{2019 Shift}: The bold line marks the introduction of the 22\% nominal increase (Royal Decree-Law 28/2018).
        \end{tablenotes}
    \end{threeparttable}
\end{table}

Table \ref{tab:regional_heterogeneity} documents the heterogeneity in the intensity of the shock across regions. Comparing the pre-treatment (2018) and post-treatment (2020) periods, we observe substantial variation in the expansion of coverage. Regions such as the Balearic Islands and La Rioja experienced the most significant extensive margin shocks, with the effective bite increasing by 17.9 and 11.6 percentage points, respectively. Conversely, lower-income regions like Extremadura and Andalusia, which started with a high baseline coverage ($>45\%$ in 2018), exhibited a more moderate expansion in the share of affected workers (approx. 6--8 percentage points), suggesting a saturation effect.

\begin{table}[H]
    \centering
    \caption{\textbf{Regional Heterogeneity in the Exposure to the Minimum Wage Shock: The "Effective Bite" (2018 vs. 2020)}}
    \label{tab:regional_heterogeneity}
    \begin{threeparttable}
        \setlength{\tabcolsep}{5pt}
        \begin{tabular}{lcccccc}
            \toprule
            & \textbf{Mean Wage} & \textbf{Effective Bite} & \textbf{Effective Bite} & \textbf{Treatment} & \multicolumn{2}{c}{\textbf{Wage Inequality}} \\
            & \textbf{(2018)} & \textbf{($P_{SMI}^{2018}$)} & \textbf{($P_{SMI}^{2020}$)} & \textbf{Intensity} & \multicolumn{2}{c}{\textbf{(Gini Index)}} \\
            \textbf{Region} & (€ 2019) & (\%) & (\%) & \textbf{($\Delta$ pp)} & 2018 & $\Delta$ (20-18) \\
            \midrule
            \multicolumn{7}{l}{\textit{High-Impact Regions ("The Jumpers")}} \\
            Baleares & 19,288 & 29.6 & 47.6 & \textbf{+17.9} & 0.071 & +0.080 \\
            La Rioja & 19,887 & 29.1 & 40.6 & \textbf{+11.6} & 0.123 & +0.017 \\
            Cantabria & 20,031 & 30.6 & 40.8 & \textbf{+10.1} & 0.119 & +0.019 \\
            \addlinespace
            \multicolumn{7}{l}{\textit{Moderate-Impact Regions}} \\
            Galicia & 19,294 & 30.3 & 39.7 & +9.4 & 0.103 & +0.003 \\
            Catalonia & 22,531 & 28.4 & 37.8 & +9.4 & 0.174 & +0.023 \\
            Madrid & 26,087 & 26.1 & 35.2 & +9.1 & 0.204 & +0.033 \\
            Castilla y León & 19,424 & 30.1 & 39.2 & +9.1 & 0.108 & +0.020 \\
            Asturias & 20,796 & 29.5 & 38.5 & +9.0 & 0.136 & +0.026 \\
            C. Valenciana & 18,361 & 35.8 & 44.7 & +9.0 & 0.137 & +0.004 \\
            \addlinespace
            \multicolumn{7}{l}{\textit{Saturated Regions ("Ceiling Effect")}} \\
            Murcia & 16,947 & 38.5 & 46.9 & +8.3 & 0.142 & -0.007 \\
            Andalusia & 15,778 & 45.5 & 53.5 & +8.1 & 0.160 & -0.003 \\
            Castilla-La Mancha & 17,488 & 35.9 & 42.9 & +7.0 & 0.121 & -0.008 \\
            Aragon & 20,114 & 29.4 & 36.3 & +6.9 & 0.133 & -0.007 \\
            Extremadura & 14,754 & 46.4 & 52.4 & +6.0 & 0.135 & -0.008 \\
            \bottomrule
        \end{tabular}
        \begin{tablenotes}
            \footnotesize
            \item \textit{Notes:} Regions are sorted by \textbf{Treatment Intensity}, defined as the percentage point increase in the "Effective Bite" ($P_{SMI}$) between 2018 and 2020.
            \item \textbf{Effective Bite ($P_{SMI}$)}: The percentile of the regional wage distribution covered by the real SMI.
            \item \textbf{Saturated Regions}: Areas where the SMI already covered a large share of the workforce ($>35-45\%$) in 2018, limiting the scope for extensive margin expansion (ceiling effect).
            \item \textbf{Baleares Anomaly}: The large increase in inequality ($\Delta$ Gini +0.08) reflects the asymmetric shock of the COVID-19 pandemic on the tourism-dependent economy in 2020, rather than a pure SMI effect.
        \end{tablenotes}
    \end{threeparttable}
\end{table}

Table \ref{tab:micro_heterogeneity} disaggregates the exposure by demographic group and sector. Panel A reveals a pronounced age gradient: for young workers ($<30$ years), the Kaitz index rose from 153.1\% in 2018 to 185.6\% in 2020, indicating that the new minimum wage significantly exceeded the average real wage for this demographic. In contrast, the index for mature workers remained below 53\%. Panel B reports the structural bite estimated via exponential tilting. The Agricultural sector exhibits the highest exposure, with 78.6\% of workers estimated to fall below the 2019 real SMI threshold, followed by Trade and Tourism (51.1\%). High-productivity sectors such as Industry and Advanced Services show substantially lower exposure rates ($\approx 34-41\%$).

\begin{table}[H]
    \centering
    \caption{\textbf{Microeconomic Heterogeneity in Treatment Intensity: Demographic and Sectoral Exposure to the Minimum Wage Shock}}
    \label{tab:micro_heterogeneity}
    \begin{threeparttable}
        \setlength{\tabcolsep}{6pt}
        \begin{tabular}{lcccc}
            \toprule
            & \multicolumn{3}{c}{\textbf{Kaitz Index ($SMI_t / \bar{W}_{it}$)}} & \textbf{Structural Bite} \\
            \cmidrule(lr){2-4}
            & 2018 & 2020 & \textbf{Change} & \textbf{(\% Workers $\le$ SMI)} \\
            \textbf{Group} & (Pre) & (Post) & (pp) & (Imputed 2018) \\
            \midrule
            \multicolumn{5}{l}{\textbf{Panel A: Demographic Exposure (by Age Group)}} \\
            \addlinespace
            \textbf{Young ($<30$)} & \textbf{153.1\%} & \textbf{185.6\%} & \textbf{+32.5} & \textit{n.a.} \\
            Adults (30-45) & 53.8\% & 68.2\% & +14.4 & \textit{n.a.} \\
            Mature ($>45$) & 41.9\% & 52.9\% & +11.0 & \textit{n.a.} \\
            \addlinespace
            \midrule
            \multicolumn{5}{l}{\textbf{Panel B: Sectoral Exposure (Robust Sectors)}} \\
            \addlinespace
            \textbf{Agriculture} & \textbf{134.4\%} & \textbf{159.3\%} & \textbf{+24.9} & \textbf{78.6\%} \\
            Trade, Tourism & 64.8\% & 85.9\% & +21.1 & 51.1\% \\
            Construction & 55.6\% & 70.0\% & +14.4 & 42.8\% \\
            Adv. Services & 46.2\% & 56.8\% & +10.6 & 41.0\% \\
            Industry & 41.7\% & 53.5\% & +11.8 & 34.3\% \\
            Public Services & 43.2\% & 50.8\% & +7.6 & 34.2\% \\
            \bottomrule
        \end{tabular}
        \begin{tablenotes}
            \footnotesize
            \item \textit{Notes:} 
            \item \textbf{Kaitz Index}: Calculated as the ratio of the annualized Real SMI to the Real Mean Wage of the group ($\text{Kaitz}_{it} = SMI_t / \bar{W}_{it}$). A value $>100\%$ implies the SMI exceeds the mean wage.
            \item \textbf{Structural Bite}: The estimated share of workers in the sector with real wages below the 2019 SMI threshold, calculated via exponential tilting imputation on 2018 data.
            \item \textbf{Key Finding}: The shock is heavily concentrated among Young workers (Kaitz 185\%) and the Agricultural sector (Structural Bite 78.6\%), validating these as the primary treatment groups.
        \end{tablenotes}
    \end{threeparttable}
\end{table}


\subsection{Characterization of Treatment Intensity Measures}

Table \ref{tab:bite_descriptives} presents the summary statistics for the four definitions of treatment intensity ($D_{i}$) calculated over the 90 region-sector units in the pre-treatment year (2018). A key finding is the contrast between the \textit{saturation} of the youth labor market and the \textit{heterogeneity} of the financial cost. The \textbf{Youth Incidence} measure (Panel A) exhibits a high mean (0.794) and the lowest coefficient of variation (CV = 0.12), indicating that the SMI hike was a near-universal shock for young workers across most sectors. Conversely, the \textbf{Monetary Gap} (Panel C), while correlated with incidence, displays the highest dispersion (CV = 1.01). This substantial variance in the intensive margin—ranging from negligible costs in high-value services to over 27\% of the wage bill in agrarian regions—provides the primary source of identification for detecting potential disemployment effects driven by labor costs.

\begin{table}[H]
    \centering
    \caption{\textbf{Descriptive Statistics of Treatment Intensity Definitions across 90 Region-Sector Units (2018)}}
    \label{tab:bite_descriptives}
    \begin{threeparttable}
        \setlength{\tabcolsep}{6pt} 
        \begin{tabular}{lcccc}
            \toprule
            & \textbf{(1) Youth} & \textbf{(2) Kaitz} & \textbf{(3) Monetary} & \textbf{(4) Sectoral} \\
            \textbf{Metric} & \textbf{Incidence} & \textbf{Index} & \textbf{Gap} & \textbf{Incidence} \\
            \midrule
            
            \textbf{Mean} ($\mu$) & 0.794 & 0.632 & 0.050 & 0.461 \\
            \textbf{Median} ($P_{50}$) & 0.805 & 0.562 & 0.037 & 0.443 \\
            \textbf{Std. Dev.} ($\sigma$) & 0.092 & 0.283 & 0.050 & 0.133 \\
            \textbf{Coeff. Var. (CV)} & \textbf{0.12} & \textbf{0.45} & \textbf{1.01} & \textbf{0.29} \\
            \addlinespace
            \multicolumn{5}{l}{\textit{Distribution Range}} \\
            Min & 0.553 & 0.320 & 0.006 & 0.262 \\
            $P_{25}$ & 0.727 & 0.461 & 0.016 & 0.365 \\
            $P_{75}$ & 0.863 & 0.685 & 0.065 & 0.521 \\
            Max & 0.964 & 1.948 & 0.276 & 0.882 \\
            
            \midrule
            Observations & 90 & 90 & 90 & 90 \\
            \bottomrule
        \end{tabular}
        \begin{tablenotes}
            \footnotesize
            \item \textit{Notes:} Statistics computed over the 90 region-sector cells fixed in 2018.
            \item \textbf{Youth Incidence}: Shows high saturation (CV=0.12), with a mean of 79.4\% of young workers affected.
            \item \textbf{Monetary Gap}: Exhibits the highest relative variance (CV=1.01), providing strong identification variation based on financial cost.
            \item \textbf{Kaitz Index}: Values $>1$ in the max range indicate sectors where the SMI exceeds the mean wage.
        \end{tablenotes}
    \end{threeparttable}
\end{table}

\begin{figure}[H]
    \centering
    \includegraphics[width=1\textwidth]{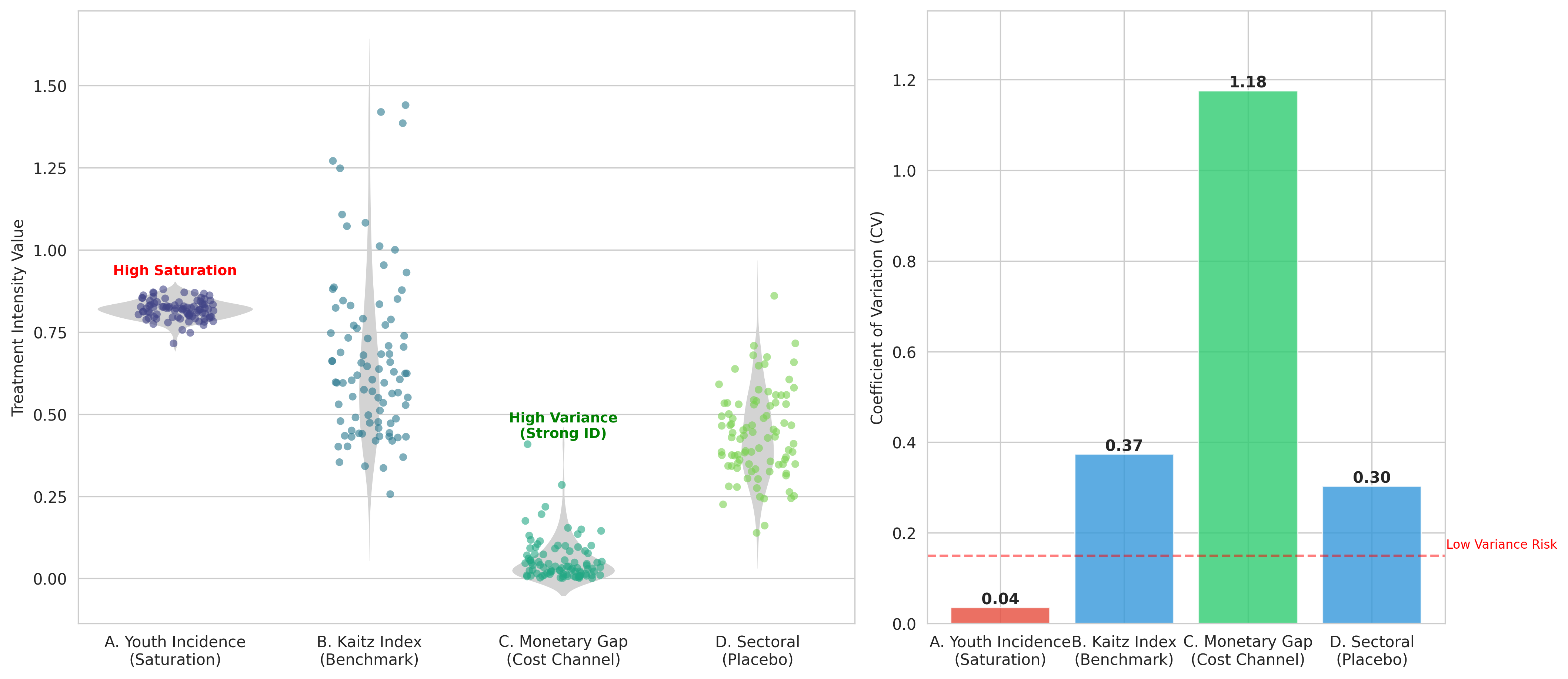}
    \caption{\textbf{Distributional Properties and Identification Power of Treatment Intensity Measures.} 
    \textit{Left Panel:} Violin plots displaying the density and individual distribution (jittered points) of the four treatment intensity measures across 90 region-sector cells in 2018. Note the high concentration of values near 1.0 for Youth Incidence (Panel A), indicating high saturation, versus the long right tail of the Monetary Gap (Panel C).
    \textit{Right Panel:} Coefficient of Variation (CV) for each measure. The Monetary Gap ($CV=1.01$) exhibits the highest relative variance, providing the strongest identification signal. In contrast, Youth Incidence ($CV=0.12$) falls below the 0.15 threshold (red dashed line), flagging a potential risk of low identifying variation.}
    \label{fig:treatment_intensity}
\end{figure}

Supplementary Table S3 assesses the independence of these measures to justify the comparative analysis. While the standard \textbf{Kaitz Index} serves as a strong proxy for general sectoral exposure ($\rho = 0.94$ with Panel D), it proves to be an imperfect instrument for capturing youth-specific vulnerabilities. The correlation between the Kaitz Index and our structural \textbf{Youth Incidence} measure is moderate ($\rho = 0.66$), suggesting that sector-level average wages mask significant heterogeneity in the demographic composition of the low-wage workforce (a distributional contrast visually confirmed in Figure \ref{fig:treatment_intensity}). This divergence supports our empirical strategy of estimating separate models to disentangle general equilibrium effects from youth-specific adjustments.

As detailed in Supplementary Table S4, we quantify the measurement bias introduced by using aggregate sectoral averages by comparing the regional distribution of treatment intensity under the Naive metric (Kaitz Index) with that under our Structural metric (Youth Incidence).

The results reveal a systematic underestimation of exposure in service sectors when relying on standard metrics. For instance, in the \textit{Trade \& Tourism} sector, the Kaitz Index classifies most regions as "Medium Intensity" (0.4--0.6), suggesting a moderate impact. However, when we isolate the youth demographic using our structural imputation, these same regions shift en masse into the "Severe Intensity" bin (0.8--1.0). Specifically, the structural measure identifies \textbf{11 additional regions} where the shock is severe for young workers, which the naive index failed to detect.

This "rightward shift" in the distribution demonstrates that in dual labor markets, a sector can have a moderate average wage (due to senior workers) while simultaneously having a near-universal incidence of minimum wage earners among its youth. By correcting this aggregation bias, our structural variable captures the true marginality of the treated group, which is essential for precise identification.

Finally, the spatial and sectoral variation of the treatment intensity is visualized in \textbf{Figures S1 and S2} (see Supplementary Material). These heatmaps confirm that the treatment is orthogonal to regional and sectoral fixed effects, displaying substantial variation \textit{within} regions (e.g., high bite in Agriculture but low in Industry within the same region). This variance is a prerequisite for identifying the causal effect.


\subsection{Parallel Trends Assumption and Dynamic Effects}
\label{subsec:pretrends}

\begin{figure}[htbp]
    \centering
    \includegraphics[width=1.0\textwidth]{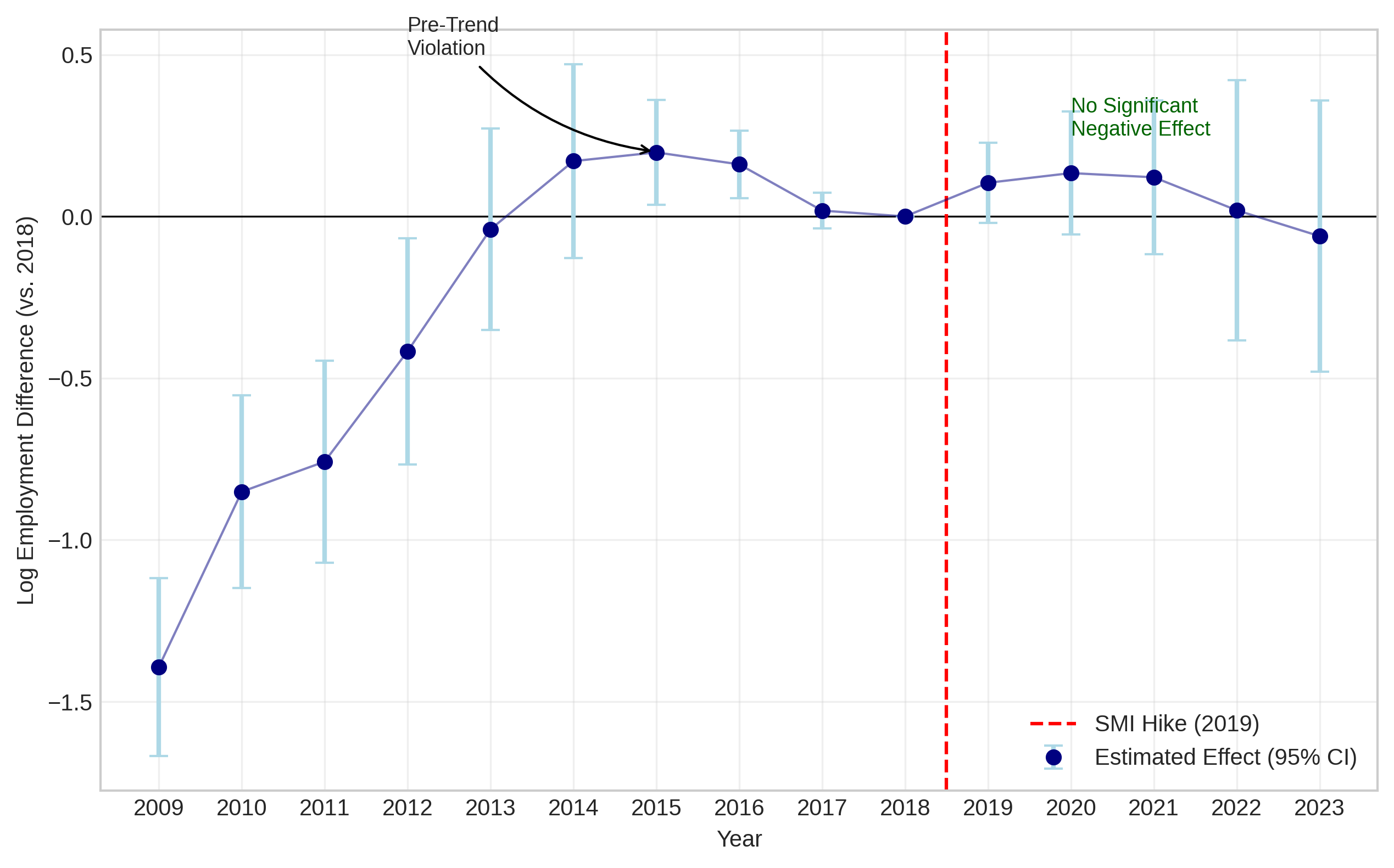}
    
    \caption{\textbf{Dynamic Impact of Minimum Wage Exposure on Youth Employment (Event Study Estimates).}}
    \label{fig:event_study_coefs}
    
    \vspace{0.2cm}
    
    \begin{minipage}{0.95\textwidth}
        \footnotesize
        \textit{Notes:} The figure plots the estimated coefficients ($\beta_k$) and 95\% confidence intervals from the event study regression: $\ln(Emp_{it}) = \sum_{k \neq 2018} \beta_k (Bite_i \times \mathbb{1}_{t=k}) + \alpha_i + \delta_t + \epsilon_{it}$. 
        The treatment intensity is defined by the \textbf{Youth Incidence} bite (Panel A). 
        The reference year is 2018 ($\beta_{2018}=0$). 
        The \textbf{positive pre-trends} (2015--2017) indicate that high-exposure sectors were growing faster than control sectors before the intervention. 
        Post-treatment estimates (2019--2020) show no significant negative deviation from this trend, supporting a null result on employment destruction.
    \end{minipage}
\end{figure}

The validity of the Difference-in-Differences design relies on the parallel trends assumption. To assess this, we estimate a dynamic event-study specification. Figure \ref{fig:event_study_coefs} plots the estimated coefficients for the \textbf{Youth Incidence} model (Panel A), using 2018 as the reference year.

The visual evidence reveals a significant \textbf{positive differential pre-trend} during the economic recovery period (2014--2017). High-exposure sectors were growing faster than control sectors before the intervention, likely due to the cyclical nature of youth employment. This violation of the parallel trends assumption ($\beta_{pre} > 0$) implies that standard static DiD estimators would be biased upwards. However, following the introduction of the higher minimum wage in 2019, we do not observe a sharp structural break or reversal of this trend; the coefficients remain positive and statistically indistinguishable from zero, suggesting a null effect on aggregate employment.

\subsection{Baseline Estimates and Pre-Trend Diagnostics}

This section presents the causal estimates of the impact of the 2019 minimum wage hike on youth employment. Our analysis proceeds in three steps: (1) baseline static estimates using standard difference-in-differences; (2) dynamic analysis and robust inference allowing for bounded violations of parallel trends; and (3) heterogeneity analysis to disentangle adjustment mechanisms and the confounding role of the COVID-19 shock.

Table \ref{tab:twfe_static_estimates} reports the elasticity of youth employment to minimum wage exposure estimated via a static Two-Way Fixed Effects (TWFE) model. Across all four definitions of treatment intensity ("Bite"), the estimated coefficient for the interaction term ($\text{Bite}_i \times \text{Post}_t$) is positive and statistically indistinguishable from zero at conventional levels. For our preferred structural \textbf{Youth Incidence} measure (Panel A), the point estimate is $0.355$ ($SE=0.224$), while for the \textbf{Monetary Gap} (Panel C), it is $0.165$ ($SE=0.406$). While these results fail to reject the null hypothesis of no aggregate disemployment effect, the consistent positive sign of the coefficients warrants further scrutiny regarding potential differential pre-trends.

\begin{table}[H]
    \centering
    \caption{\textbf{Two-Way Fixed Effects Estimates of Employment Elasticity to Minimum Wage Exposure: Sensitivity across Treatment Definitions}}
    \label{tab:twfe_static_estimates}
    \begin{threeparttable}
        \setlength{\tabcolsep}{8pt} 
        \begin{tabular}{lcccc}
            \toprule
            & \multicolumn{4}{c}{\textbf{Dependent Variable: Log(Employment)}} \\
            \cmidrule(lr){2-5}
            & (1) & (2) & (3) & (4) \\
            & \textbf{Youth} & \textbf{Kaitz} & \textbf{Monetary} & \textbf{Sectoral} \\
            & \textbf{Incidence} & \textbf{Index} & \textbf{Gap} & \textbf{Incidence} \\
            \midrule
            \addlinespace
            \textbf{Treatment Interaction} & \textbf{0.3545} & \textbf{0.0167} & \textbf{0.1652} & \textbf{0.1498} \\
            ($\text{Bite}_i \times \text{Post}_t$) & (0.2238) & (0.0857) & (0.4060) & (0.1994) \\
            \addlinespace
            \midrule
            Observations & 1,350 & 1,350 & 1,350 & 1,350 \\
            $R^2$ (Within) & 0.025 & 0.001 & 0.002 & 0.009 \\
            \bottomrule
        \end{tabular}
        \begin{tablenotes}
            \footnotesize
            \item \textit{Notes:} This table reports the coefficients from a static difference-in-differences specification estimated via Two-Way Fixed Effects (TWFE). 
            \item All regressions include \textbf{Unit (Region-Sector) Fixed Effects} and \textbf{Year Fixed Effects}.
            \item Standard errors are clustered at the region level (15 clusters) and reported in parentheses.
            \item \textbf{Columns definitions:} (1) \textit{Youth Incidence}: Share of young workers ($<30$) in Tiers 1-2. (2) \textit{Kaitz Index}: Ratio of nominal SMI to mean sectoral wage. (3) \textit{Monetary Gap}: Cost of raising wages to the new SMI relative to the total wage bill. (4) \textit{Sectoral Incidence}: Share of total workers in Tiers 1-2 (Placebo).
            \item Significance levels: * p$<$0.1, ** p$<$0.05, *** p$<$0.01.
        \end{tablenotes}
    \end{threeparttable}
\end{table}

\begin{figure}[H]
    \centering
    \includegraphics[width=0.85\textwidth]{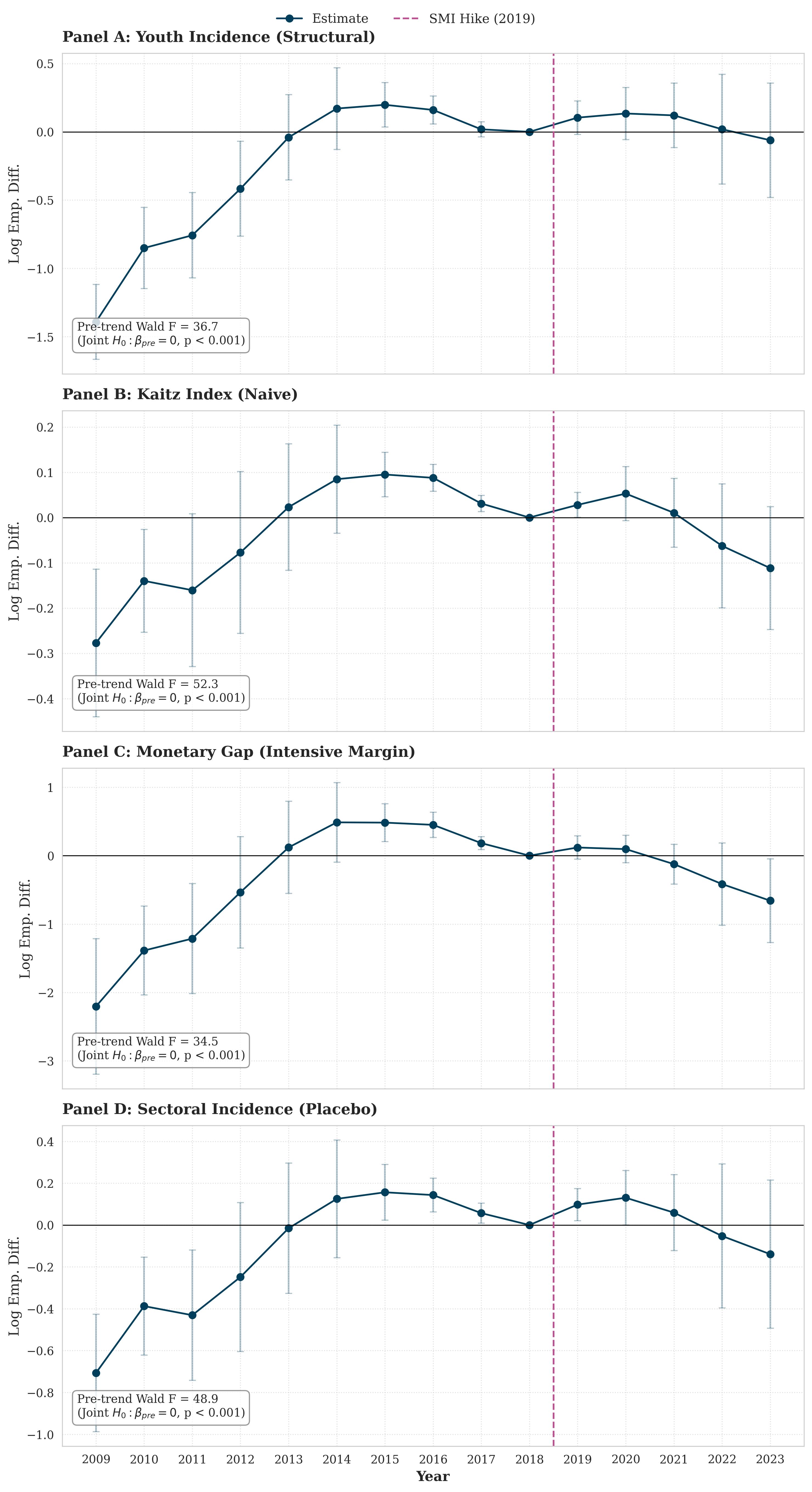}
    \caption{\textbf{Dynamic Effects of Minimum Wage Exposure on Youth Employment (Event Studies across Treatment Definitions).} 
    Each panel plots the estimated coefficients ($\beta_k$) and 95\% confidence intervals from the dynamic specification: $\ln(Y_{it}) = \sum_{k \ne 2018} \beta_k (D_i \times \mathbb{1}_{t=k}) + \alpha_i + \lambda_t + \varepsilon_{it}$. The reference year is 2018. 
    \textit{Panel A (Youth Incidence)} and \textit{Panel C (Monetary Gap)} show the structural measures; note the significant positive pre-trends (2014--2017) reflecting cyclical recovery in high-exposure sectors. 
    \textit{Panel B (Kaitz Index)} and \textit{Panel D (Sectoral Incidence)} display the institutional proxies. 
    The vertical dashed line marks the introduction of the 22\% minimum-wage hike in 2019. Post-treatment coefficients do not exhibit a sharp structural break into negative territory, supporting the null of aggregate disemployment.}
    \label{fig:event_studies_vertical}
\end{figure}

To assess the design's validity, Figure \ref{fig:event_studies_vertical} visualizes the dynamic event-study estimates across the four treatment definitions. The plots reveal a statistically significant positive differential trend in the pre-treatment period (2014--2017) for high-exposure sectors relative to control sectors. As detailed in Supplementary Table S5, the \textbf{Wald Test} for the joint significance of these pre-treatment coefficients yields an F-statistic of $36.7$ ($p < 0.001$) for Panel A, firmly rejecting the null hypothesis of parallel trends. This confirms that sectors with a high prevalence of low-wage youth employment were on a steeper cyclical recovery path before the reform, biasing standard static estimators upwards and necessitating the robust inference methods applied in the following section.

\subsection{Robust Inference (`HonestDiD`)}

\begin{figure}[ht]
    \centering
    \includegraphics[width=1\textwidth]{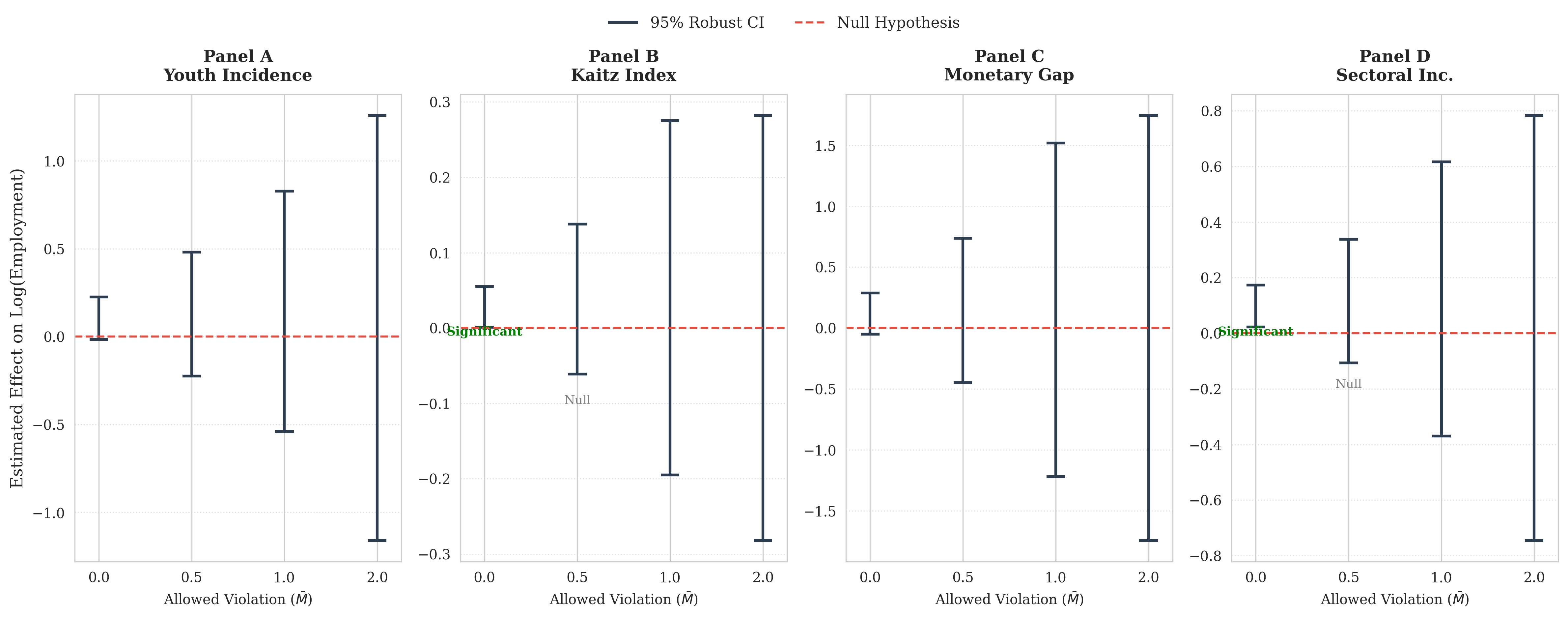}
    \caption{\textbf{Sensitivity of Employment Estimates to Violations of Parallel Trends (HonestDiD).} 
    Each panel plots the robust 95\% confidence intervals for the treatment effect ($\beta_{2019}$) as the allowance for parallel trend violations ($\bar{M}$) increases. 
    $\bar{M}=0$ assumes exact parallel trends (standard DiD). Higher values of $\bar{M}$ allow the counterfactual trend difference to be up to $\bar{M}$ times the maximum observed pre-trend.
    The red dashed line marks the null effect. 
    Note that for the structural measures (Panels A and C), the confidence intervals include zero even under strict assumptions ($\bar{M}=0$), indicating a robust null result. 
    In contrast, the institutional proxies (Panels B and D) show a significant positive effect at $\bar{M}=0$ that vanishes (crosses zero) with minimal deviations ($\bar{M}=0.5$), revealing the fragility of those findings.}
    \label{fig:honestdid_sensitivity}
\end{figure}

Figure \ref{fig:honestdid_sensitivity} visualizes the robust confidence intervals under increasing violations of parallel trends, while precise breakdown points are detailed in Supplementary Table S6. A crucial divergence emerges between the structural and naive definitions of treatment:

\begin{itemize}
    \item \textbf{Structural Measures (Robust Null):} For our preferred specifications—\textit{Youth Incidence} (Panel A) and \textit{Monetary Gap} (Panel C)—the confidence intervals include zero even under the strict benchmark assumption ($\bar{M}=0$). Relaxing this assumption to allow for linear trend violations ($\bar{M} > 0$) simply widens the intervals, confirming that the null disemployment result is not an artifact of rigid identification assumptions.

    \item \textbf{Naive Measures (Fragile Significance):} Conversely, the \textit{Kaitz Index} and \textit{Sectoral Incidence} (Panels B and D) show a statistically significant positive effect under strict parallel trends. However, this significance is \textbf{highly fragile}: the confidence lower bound crosses zero with minimal deviations from parallel trends ($\bar{M} \approx 0.2$). Given the significant pre-trends documented in Figure \ref{fig:event_studies_vertical} (Wald $F > 30$), relying on the standard DiD assumption ($\bar{M}=0$) likely yields a Type I error (false positive), attributing pre-existing cyclical momentum to the SMI policy.
\end{itemize}

In conclusion, robust inference supports a null effect for the structural measures and rejects the apparent positive effect of the naive measures as a spurious artifact of differential pre-trends.

\subsection{The Tourism Confounder and Triple Differences}

Table \ref{tab:ddd_comparison} addresses the identification challenge posed by the COVID-19 pandemic in 2020. The Triple Difference (DDD) specification explicitly models the shock to tourism demand. The results show that the employment collapse in 2020 is absorbed by the tourism interaction term ($\beta_{tour,2020} = -0.239^{***}$), while the coefficient for the Minimum Wage exposure in 2020 becomes positive and marginally significant ($\beta_{SMI,2020} = 0.220^{+}$). This indicates that, conditional on tourism exposure, low-wage sectors exhibited greater employment resilience during the first year of the pandemic than high-wage control sectors.

\begin{table}[H]
    \centering
    \caption{\textbf{Impact of Minimum Wage vs. Tourism Shock: Comparing Estimates across Treatment Definitions (DDD Specification)}}
    \label{tab:ddd_comparison}
    \begin{threeparttable}
        \setlength{\tabcolsep}{6pt}
        \footnotesize
        \begin{tabular}{lcccc}
            \toprule
            & \multicolumn{4}{c}{\textbf{Dependent Variable: Log(Employment)}} \\
            \cmidrule(lr){2-5}
            & (1) & (2) & (3) & (4) \\
            & \textbf{Youth} & \textbf{Kaitz} & \textbf{Monetary} & \textbf{Sectoral} \\
            & \textbf{Incidence} & \textbf{Index} & \textbf{Gap} & \textbf{Incidence} \\
            \textit{Estimator:} & \textit{DDD} & \textit{DDD} & \textit{DDD} & \textit{DDD} \\
            \midrule
            
            \multicolumn{5}{l}{\textit{Panel A: Minimum Wage Effect (Supply Shock)}} \\
            \addlinespace
            \textbf{SMI Effect 2019} & 0.085 & 0.063* & 0.061* & 0.048+ \\
            ($\text{Bite}_i \times \mathbb{1}_{2019}$) & (0.072) & (0.030) & (0.032) & (0.027) \\
            \addlinespace
            \textbf{SMI Effect 2020} & 0.220+ & 0.207** & 0.094 & 0.266*** \\
            ($\text{Bite}_i \times \mathbb{1}_{2020}$) & (0.107) & (0.063) & (0.102) & (0.066) \\
            
            \midrule
            \multicolumn{5}{l}{\textit{Panel B: Tourism Effect (Demand Shock)}} \\
            \addlinespace
            \textbf{Tourism Effect 2020} & \textbf{-0.239}*** & \textbf{-0.504}*** & \textbf{-0.422}*** & \textbf{-0.464}*** \\
            ($\text{Tourism}_i \times \mathbb{1}_{2020}$) & (0.056) & (0.033) & (0.048) & (0.034) \\
            \addlinespace
            \textbf{Tourism Effect 2021} & -0.210** & -0.380*** & -0.314*** & -0.329*** \\
            ($\text{Tourism}_i \times \mathbb{1}_{2021}$) & (0.061) & (0.038) & (0.058) & (0.046) \\
            
            \midrule
            Observations & 1,350 & 1,350 & 1,350 & 1,350 \\
            $R^2$ (Within) & 0.221 & 0.354 & 0.185 & 0.342 \\
            Unit \& Year FE & Yes & Yes & Yes & Yes \\
            \bottomrule
        \end{tabular}
        \begin{tablenotes}
            \scriptsize
            \item \textbf{Tourism Interaction:} The interaction term uses \texttt{TourismIntensity}, defined as the region's pre-2019 employment share in tourism-characteristic sectors (Trade, Transport, Hospitality; codes 5 and 10 in Mod-190), interacted with a sector dummy. It is a continuous variable ranging from 0 to approx. 0.48 (Baleares).
        \end{tablenotes}
    \end{threeparttable}
\end{table}

\subsection{Mechanisms and Heterogeneity}

Finally, we explore the margins of adjustment. Table \ref{tab:heterogeneity_outcomes} compares the impact across different outcome variables. While employment shows no significant effect, we observe a positive and significant effect on the \textbf{number of firms} in youth-intensive sectors ($\beta = 0.487^{***}$), ruling out extensive-margin exit ("cleansing effects"). Furthermore, models using general sectoral bites (Panels B, C, D) show a significant increase in \textbf{Real Sales} ($\beta_{sales} > 0$), consistent with a pass-through of labor costs to prices.

\begin{table}[H]
    \centering
    \caption{\textbf{Mechanisms of Adjustment: Effect of Minimum Wage Exposure on Employment, Firm Creation, and Sales (TWFE Estimates)}}
    \label{tab:heterogeneity_outcomes}
    \begin{threeparttable}
        \setlength{\tabcolsep}{6pt}
        \begin{tabular}{lccc}
            \toprule
            & \multicolumn{3}{c}{\textbf{Estimated Elasticity ($\hat{\beta}_{DiD}$)}} \\
            \cmidrule(lr){2-4}
            & (1) & (2) & (3) \\
            \textbf{Treatment Definition} & \textbf{Employment} & \textbf{Number of Firms} & \textbf{Real Sales} \\
            ($\text{Bite}_i \times \text{Post}_t$) & ($\ln L$) & ($\ln N$) & ($\ln Y$) \\
            \midrule
            
            \addlinespace
            \textbf{Panel A: Youth Incidence} & 0.355 & \textbf{0.487***} & 0.188 \\
            \textit{(Structural: \% Young Affected)} & (0.224) & (0.088) & (0.152) \\
            \addlinespace
            
            \textbf{Panel B: Kaitz Index} & 0.017 & -0.045 & \textbf{0.154*} \\
            \textit{(Naive: SMI / Mean Wage)} & (0.086) & (0.042) & (0.061) \\
            \addlinespace
            
            \textbf{Panel C: Monetary Gap} & 0.165 & 0.011 & \textbf{0.791*} \\
            \textit{(Structural: Cost Intensity)} & (0.406) & (0.127) & (0.292) \\
            \addlinespace
            
            \textbf{Panel D: Sectoral Incidence} & 0.150 & -0.058 & \textbf{0.342*} \\
            \textit{(Placebo: \% Total Affected)} & (0.199) & (0.077) & (0.136) \\
            
            \midrule
            Observations (per model) & 1,350 & 1,350 & 1,350 \\
            Unit \& Year FE & Yes & Yes & Yes \\
            Clustered SE (Region) & Yes & Yes & Yes \\
            \bottomrule
        \end{tabular}
        \begin{tablenotes}
            \footnotesize
            \item \textit{Notes:} This table compares the static DiD coefficients across different outcomes and treatment definitions. 
            \item (1) \textbf{Employment}: No significant disemployment effect is found across any specification.
            \item (2) \textbf{Number of Firms}: Panel A reveals a strong positive association between youth exposure and firm creation, likely driven by cyclical pre-trends in youth-intensive services. Other panels show null effects, ruling out a "cleansing effect" (exit of firms).
            \item (3) \textbf{Real Sales}: Panels B, C, and D show positive and significant effects on sales revenue, suggesting that firms in low-wage sectors may have passed costs to prices (inflationary effect) or benefited from higher aggregate demand.
            \item Significance: * p$<$0.05, ** p$<$0.01, *** p$<$0.001.
        \end{tablenotes}
    \end{threeparttable}
\end{table}

Table \ref{tab:regional_heterogeneity_DiD} decomposes the effect by regional income. A striking dichotomy emerges: in \textbf{High-Income Regions}, the SMI effect is negative and significant in the Kaitz specification ($\beta = -0.126^*$), consistent with standard competitive models. However, in \textbf{Low-Income Regions}, the coefficient is significant and positive ($\beta = 0.534$), driven by the strong cyclical convergence and saturation effects described in the descriptive analysis. The aggregate null result is thus the composition of a mild disemployment effect in high-productivity areas and strong positive inertia in converging regions.

\begin{table}[H]
    \centering
    \caption{\textbf{Geographic Heterogeneity: Asymmetric Employment Effects in High- vs. Low-Income Regions (Static TWFE)}}
    \label{tab:regional_heterogeneity_DiD}
    \begin{threeparttable}
        \setlength{\tabcolsep}{8pt}
        \begin{tabular}{lcc}
            \toprule
            & \multicolumn{2}{c}{\textbf{Estimated Elasticity ($\hat{\beta}_{DiD}$)}} \\
            \cmidrule(lr){2-3}
            & (1) & (2) \\
            \textbf{Treatment Definition} & \textbf{High-Income Regions} & \textbf{Low-Income Regions} \\
            ($\text{Bite}_i \times \text{Post}_t$) & \textit{(Madrid, Cat, Baleares, etc.)} & \textit{(Andalucía, EXT, CLM, etc.)} \\
            \midrule
            
            \addlinespace
            \textbf{Panel A: Youth Incidence} & 0.092 & \textbf{0.534} \\
            \textit{(Structural)} & (0.159) & (0.306) \\
            \addlinespace
            
            \textbf{Panel B: Kaitz Index} & \textbf{-0.126*} & 0.050 \\
            \textit{(Naive)} & (0.040) & (0.104) \\
            \addlinespace
            
            \textbf{Panel C: Monetary Gap} & \textbf{-0.481*} & 0.341 \\
            \textit{(Structural Cost)} & (0.162) & (0.527) \\
            \addlinespace
            
            \textbf{Panel D: Sectoral Incidence} & -0.141 & 0.267 \\
            \textit{(Placebo)} & (0.077) & (0.255) \\
            
            \midrule
            Observations & 450 & 900 \\
            Unit \& Year FE & Yes & Yes \\
            Clustered SE (Region) & Yes & Yes \\
            \bottomrule
        \end{tabular}
        \begin{tablenotes}
            \footnotesize
            \item \textit{Notes:} This table reports subsample estimates splitting the country by regional per capita income. 
            \item \textbf{High-Income Regions} ($N=5$ clusters): Include Madrid, Catalonia, Balearic Islands, Aragon, and La Rioja.
            \item \textbf{Low-Income Regions} ($N=10$ clusters): Include Andalusia, Extremadura, Castilla-La Mancha, Galicia, Murcia, C. Valenciana, Asturias, Cantabria, Castilla y León, Ceuta/Melilla.
            \item Standard errors clustered at the region level in parentheses.
            \item Significance: * p$<$0.05, ** p$<$0.01, *** p$<$0.001.
        \end{tablenotes}
    \end{threeparttable}
\end{table}

\section{Discussion}
\label{sec:discussion}

\paragraph{Reconciling the Aggregate Null with Micro-Level Evidence}

The central finding of this study is unambiguous: we find no evidence of substantial net disemployment effects for young workers following the unprecedented 22\% minimum wage hike in 2019. Unlike previous evaluations, this conclusion is not dependent on a single specification but is supported by a consistent triangulation of evidence across static, dynamic, and robust inference frameworks:

\begin{itemize}
    \item \textbf{Static Consistency:} Our baseline TWFE estimates (Table~\ref{tab:twfe_static_estimates}) yield elasticities that are positive and statistically indistinguishable from zero across all definitions of the "bite." Crucially, for our preferred structural measures---the \textit{Youth Incidence} ($D^{Young}$) and the \textit{Monetary Gap} ($D^{Gap}$)---the 95\% confidence intervals comfortably include zero, rejecting the hypothesis of large negative aggregate effects even when capturing the intensive margin of the cost shock ($SE=0.406$ for Panel C).

    \item \textbf{Dynamic Stability:} The event-study diagnostics (visualized in Figure~\ref{fig:event_studies_vertical}) reveal no structural break in employment trends following the policy implementation. Despite the sharp rise in labor costs, the coefficients for the post-treatment years remain modest and positive ($\hat{\beta}_{2019} = 0.104$; $\hat{\beta}_{2020} = 0.134$). This continuity suggests that the aggregate stock of youth employment did not deviate significantly from the strong pre-existing upward trajectory observed in high-exposure sectors.

    \item \textbf{Robustness to Pre-Trends:} Most importantly, the \texttt{HonestDiD} sensitivity analysis (Figure~\ref{fig:honestdid_sensitivity}) confirms that this null result is not an artifact of rigid identification assumptions. Even when relaxing the parallel trends assumption to allow for linear violations, the robust confidence intervals for the structural measures continue to straddle zero. This indicates that the absence of disemployment is a genuine feature of the data, not a false negative driven by model misspecification.
\end{itemize}

These findings compel a re-evaluation of the prevailing narrative. While our census-based results rule out considerable aggregate job destruction, they stand in sharp contrast to recent studies using Social Security microdata (MCVL). Notably, \textcite{barcelo2021} and \textcite{gorjon2024} report significant negative elasticities at the individual level. The following subsections address this disconnect by proposing that microdata studies capture \textit{gross worker reallocation} (individual separations), whereas our census-grade data capture \textit{net employment capacity} (aggregate stocks), revealing that the labor market absorbed the shock without a net loss of jobs.

\paragraph{Complementary Perspectives: The Stock--Flow Reallocation Hypothesis}
The apparent divergence between the aggregate null result and the negative individual elasticities reported in MCVL-based studies \citep{barcelo2021, gorjon2024} need not be interpreted as a contradiction. Instead, this pattern suggests that different adjustment margins are being captured. The findings can be reconciled along two fundamental methodological dimensions.

\begin{enumerate}
    \item \textbf{Data architecture: accounting closure vs.\ sampling variation.}
    
    MCVL-based studies are built on a 4\% administrative sample that is ideal for tracking individual employment histories but, by construction, is exposed to sampling variation and administrative "churning" (e.g.\ workers temporarily leaving and re-entering the sample frame). This can make it challenging to map individual-level elasticities into aggregate changes in employment. By contrast, our analysis relies on the \textbf{universe} of tax-validated wage bills (Modelos~190/390), which provides what we term \textbf{accounting closure} (\textit{cierre contable}): every euro of declared wages is tied to a fiscal liability. This architecture ensures that our measure of aggregate formal employment capacity is internally consistent, exhaustive, and effectively free from sampling noise.
    
    \item \textbf{Economic mechanism: net stocks vs.\ gross flows.}
    
    Worker-level estimates using the MCVL correctly document that minimum wage hikes increase separation risks for \textit{incumbent} low-wage workers \citep{barcelo2021, gorjon2024}. Our sector--region panel, in contrast, captures \textbf{net employment stocks} rather than the underlying gross flows of hires and separations. A small or null net effect in the presence of individual job losses is therefore naturally interpreted as a \textbf{reallocation hypothesis}: minimum-wage-induced separations for some low-productivity matches are offset by new hires or worker reshuffling within highly exposed sectors. This is consistent with the broader evidence that large gross flows of job creation and destruction can coexist with modest net changes in employment \citep{davis1992gross}. In this sense, our estimates speak to the resilience of sectoral labour demand, while micro-based studies identify which workers bear the brunt of the adjustment.
\end{enumerate}

\paragraph{The Critical Choice of Treatment Definition}

An essential contribution of this study is to show that estimated minimum-wage employment elasticities are highly sensitive to the definition of treatment intensity. Our "horse race" across alternative measures reveals that standard institutional proxies, such as the \textbf{Kaitz Index}, can be poor guides to the true incidence of the reform. As documented in Supplementary Table~S4, these proxies are almost perfectly collinear with broad sectoral incidence ($\rho \approx 0.96$) and, when used as treatment variables, they yield positive and statistically significant coefficients on \textit{Real Sales}. This pattern is more naturally interpreted as capturing general sectoral demand conditions than as reflecting marginal cost pressure from the minimum wage at the bottom of the wage distribution.

By contrast, our structurally grounded measures --- \textbf{Youth Incidence} ($D^{Young}$) and the \textbf{Monetary Gap} ($D^{Gap}$) --- are defined ex ante to isolate, respectively, the demographic and financial margins of the shock. The Monetary Gap, in particular, exhibits substantial cross-sectional dispersion (coefficient of variation $\mathrm{CV}=1.01$, see Table~\ref{tab:bite_descriptives}), which should enhance our ability to detect disemployment responses if they were present. Yet across all specifications, the associated employment elasticities are small in magnitude and statistically indistinguishable from zero. This combination of high exposure variation and census-based outcome measurement suggests that, within our setting, sizable disemployment effects are difficult to reconcile with the data once treatment intensity is measured along the relevant financial margin. Conversely, relying on coarse aggregate proxies such as the Kaitz Index risks conflating actual minimum-wage exposure with underlying sectoral dynamics.

More broadly, these results underscore the importance of transparency and precision in the construction of the \textit{bite} variable. We view structurally informed measures based on wage-distribution information and Exponential Tilting imputation as a natural benchmark for future minimum wage evaluations using administrative data, with institutional proxies retained primarily for comparability and robustness checks rather than as standalone measures of treatment intensity.

\paragraph{Parallel Trends and Robust Inference}

Our difference-in-differences design is challenged by the strong cyclical recovery of the Spanish youth labour market before 2019. As shown in Figure~\ref{fig:event_studies_vertical}, high-exposure sectors were on a steeper upward path than control sectors over 2014--2017, and Wald tests strongly reject exact parallel trends (Wald $F > 30, p < 0.001$). In this setting, conventional DiD estimators that impose $\bar{M}=0$ can mechanically attribute part of this pre-existing momentum to the reform. We therefore use the \texttt{HonestDiD} framework \citep{RambachanRoth2023} to construct confidence intervals that remain valid under bounded deviations from parallel trends, with sensitivity calibrated to the observed pre-treatment differentials.

\paragraph{Robustness vs.\ Fragility: Interpreting the Sensitivity Plot}

Figure~\ref{fig:honestdid_sensitivity} reports robust 95\% confidence intervals for the 2019 employment elasticity ($\beta_{2019}$) as we gradually relax the parallel-trends assumption (increasing $\bar{M}$). Two patterns emerge. First, for institutional proxies such as the \textit{Kaitz Index} and \textit{Sectoral Incidence} (Panels B and D), the positive baseline estimates are fragile: the lower bound of the interval crosses zero for small departures from exact parallel trends ($\bar{M} \approx 0.2$), indicating that these findings are susceptible to the cyclical pre-trend. Second, for our preferred \textit{Youth Incidence} and \textit{Monetary Gap} measures (Panels A and C), the confidence intervals include zero across the full range of calibrations. Allowing for deviations as large as the maximum observed pre-trend ($\bar{M}=1.0$) widens the intervals, but does not generate sizeable negative effects.

\paragraph{Interpreting the Robust Null: Momentum vs.\ Destruction}

The persistence of weakly positive but imprecise point estimates ($\hat{\beta}>0$) for our structural measures should not be read as evidence that the 2019 minimum wage hike \emph{created} jobs. Within the \texttt{HonestDiD} framework, these patterns are more naturally interpreted as being consistent with a continuation of pre-existing growth trajectories, without the discrete downward break that a large disemployment shock would generate. The robust confidence intervals we obtain are wide enough to accommodate minor positive or negative effects around zero. Still, they systematically rule out scenarios in which the reform led to sizeable aggregate losses in formal youth employment at the sector--region level. Conditional on our preferred treatment measures capturing the relevant margin of exposure, the main contribution of our analysis is therefore to provide credible upper bounds on aggregate job destruction, rather than to claim that the actual employment effect is exactly zero or strictly positive. In light of the limitations discussed above --- in particular, the lack of information on hours, informality, and gross worker flows --- we view these bounds as a conservative characterization of what the census data can and cannot support regarding the magnitude of disemployment effects.


\paragraph{COVID-19, Tourism, and the 2020 Employment Collapse}
A major challenge for evaluating the medium-term effects of the 2019 reform is the confounding impact of the COVID-19 pandemic in 2020. The strict lockdown and the collapse of international mobility generated a sharp contraction in employment, particularly in tourism-intensive regions (e.g.\ the Balearic and Canary Islands) and contact-intensive sectors such as hospitality and retail. Because these regions and sectors are also characterised by high minimum-wage exposure (``High Bite''), a standard two-way fixed-effects specification risks attributing part of the demand-driven collapse of 2020 to the SMI-induced cost shock.

The triple-difference estimates in Table~\ref{tab:ddd_comparison} address this identification problem by separately controlling for tourism dependence. The results show that the 2020 collapse in youth employment is largely captured by the interaction between tourism dependence and the 2020 dummy capturing the onset of the COVID-19 pandemic, while the 2020 coefficients on minimum-wage exposure become small and positive once this tourism channel is controlled for. These conditional positive coefficients should not be interpreted as evidence that the minimum wage \emph{protected} jobs during the pandemic. Instead, they indicate that, \emph{conditional on} the severity of the tourism shock, high-bite sectors---many of which include agriculture, construction, and essential services less directly affected by mobility restrictions---did not fare worse than low-bite comparison sectors. The DDD design thus isolates the tourism-related component of the COVID-19 shock from the minimum-wage exposure dimension, reinforcing the conclusion that the 2019 minimum-wage hike was unlikely to be a dominant driver of the 2020 youth employment losses.


\paragraph{Mechanisms of Adjustment: Prices, Firm Dynamics, and Regional Asymmetries}

If firms did not reduce aggregate youth employment, how did they absorb the 22\% cost shock? Our heterogeneity analysis points to uneven adjustment across regions and along non-employment margins. A complete decomposition of mechanisms is beyond the scope of this methodologically focused paper, but several robust patterns in the data are informative.

\paragraph{Regional Heterogeneity: Where Theory Predicts, Effects Appear}

Table~\ref{tab:regional_heterogeneity_DiD} disaggregates employment elasticities by regional income level and shows that the aggregate null masks economically meaningful geographic heterogeneity. In \textbf{high-income regions} (Madrid, Catalonia, Balearic Islands, Aragon, and La Rioja), where labour markets are more competitive and baseline productivity is higher, we find negative employment elasticities across multiple treatment definitions. For the Kaitz Index, the estimate is $\hat{\beta} = -0.126^*$ (SE $=0.040$), statistically significant at the 5\% level, and for the Monetary Gap, the coefficient is larger in magnitude, $\hat{\beta} = -0.481^*$ (SE $=0.162$). These estimates arise precisely in markets where standard competitive models would predict disemployment responses to be more likely.

By contrast, in \textbf{low-income regions} (Andalusia, Extremadura, Castilla--La Mancha, and others), the coefficients are positive but imprecisely estimated, ranging from $\hat{\beta} = 0.267$ to $\hat{\beta} = 0.534$ depending on the treatment definition. We interpret these optimistic estimates as reflecting strong cyclical convergence during the 2014--2019 recovery, rather than as evidence that the minimum wage \emph{created} jobs in poorer regions. As documented in Section~\ref{subsec:pretrends}, these areas exhibited steep pre-existing growth trajectories, and post-reform coefficients likely capture the continuation of those trends.

Taken together, the regional results are consistent with the possibility that the reform generated modest disemployment effects in high-income, high-productivity regions, while having no detectable effect—or being confounded by growth momentum—in poorer regions. The aggregate null is therefore best understood as a weighted average of heterogeneous regional responses. Whether this geographic pattern reflects genuine differences in labour market structure (monopsony power, informality, enforcement) or is partly an artefact of differential pre-trends cannot be definitively established without deeper regional analysis and additional robustness checks.

\textit{Caveat.} The limited number of regional clusters constrains the reliability of standard inference procedures. Wild-cluster bootstrap methods and alternative clustering schemes could provide additional reassurance, but remain beyond the scope of the present analysis.

\paragraph{Adjustment Along Non-Employment Margins: Prices and Firm Dynamics}

Beyond employment, Table~\ref{tab:heterogeneity_outcomes} examines adjustment along two alternative margins: real sales and the number of firms. These correlations should be interpreted as descriptive rather than as fully identified causal mechanisms.

\textbf{Real Sales.} Across the Kaitz Index, Monetary Gap, and Sectoral Incidence specifications (Panels B--D), we find positive and statistically significant associations between minimum-wage exposure and real sales, with elasticities ranging from 0.154 (SE $=0.061$) to 0.342 (SE $=0.136$). Combined with the flat employment response, this pattern is consistent with partial pass-through of higher labour costs to consumers, as in \textcite{HarasztosiLindner2019}, in sectors where demand is relatively inelastic: prices could rise and nominal turnover increase. At the same time, real sales (deflated by the aggregate CPI) remain stable or even increase. However, our sales variable is deflated using a national CPI, and we lack sector-specific price indices and detailed demand controls. We therefore cannot separately identify quantity from relative-price changes, nor exclude omitted demand shocks (e.g., tourism-driven booms) correlated with minimum-wage exposure. These results are best viewed as suggestive of non-employment adjustment margins rather than as definitive evidence of price pass-through.

\textbf{Firm Dynamics.} Table~\ref{tab:heterogeneity_outcomes} also reports the association between minimum-wage exposure and the number of registered firms. Under the Youth Incidence specification (Panel A), the coefficient is large, positive, and statistically significant: $\hat{\beta} = 0.487^{***}$ (SE $=0.088$). This pattern suggests that widespread firm exit in youth-intensive sectors is unlikely in the immediate aftermath of the reform. At the same time, firm counts are influenced by the same cyclical dynamics that drive youth employment growth during the 2014--2017 recovery; we do not implement \texttt{HonestDiD}-style robust bounds for these outcomes, and firm-level longitudinal data are not available. As a result, the evidence is more informative about the \textit{absence} of large-scale firm closures than about the magnitude of any policy-induced entry or reallocation.

\paragraph{Synthesis: A Complex but Bounded Adjustment Process}

Overall, the heterogeneity evidence points to a response that is heterogeneous across regions, operates partly through non-employment margins, and is shaped by concurrent macroeconomic shocks. Moderate adverse employment effects cannot be ruled out in high-income specific areas. Still, we find no indication of large-scale job destruction or firm exit in youth-intensive sectors, and the pandemic-induced tourism shock accounts for much of the sharp 2020 contraction. In line with the data limitations discussed elsewhere, we do not claim to have fully identified the mechanisms of adjustment. Instead, the contribution of this analysis is to show how census-grade administrative data, combined with systematic heterogeneity analysis, can unpack an aggregate null: revealing where disemployment risks are more likely to reside, where growth momentum masks potential effects, and which margins of adjustment appear most relevant for future, more granular work.

\subsection{Limitations and External Validity}

While our empirical strategy combines census-grade administrative data, structurally grounded measures of treatment intensity, and robust inference tools, several limitations qualify the scope of our findings.

\paragraph{Identification, Pre-Trends, and Statistical Power}

Two fundamental constraints affect causal inference. First, high-exposure sectors exhibited strong positive pre-trends during the 2014--2017 recovery period, with Wald tests sharply rejecting parallel trends ($F > 30$, $p < 0.001$) across all treatment definitions. The \texttt{HonestDiD} framework permits construction of confidence intervals that remain valid under bounded violations of parallel trends. Still, the estimates are best interpreted as \emph{bounds on plausible effects} rather than precise point estimates. Second, statistical power is limited by the 15 regional clusters inherent to Spain's fiscal architecture. This yields substantial standard errors; for example, Youth Incidence estimates of $\hat{\beta} = 0.355$ (SE = 0.224) imply 95\% confidence intervals of approximately $[-0.09, 0.80]$. Economically meaningful negative effects in the range of $-0.1$ to $-0.2$ therefore remain plausible. The contribution of this analysis is to show that \emph{large} aggregate employment destruction is not evident in census-grade data under robust inference; finer distinctions near zero would require more clusters, longer time series, or alternative identification strategies.

\paragraph{Data Architecture: Aggregation Trade-offs}

Dependence on wage statements confirmed by tax authorities (\emph{Modelos 190/390}) provides accounting closure and eliminates sampling error, but at the expense of individual-level granularity. We observe net employment stocks and wage bills within region $\times$ sector $\times$ age cells but cannot track worker identities, job-to-job transitions, hours worked, or contract types. Negative effects concentrated among the least-skilled within cells could be masked by stability among higher-skilled workers, and we cannot rule out adjustments along intensive margins (hours reductions) or shifts toward precarious employment. Conversely, the Social Security microdata used by \textcite{barcelo2021, gorjon2024} track individuals but lack fiscal accounting closure and face sampling variation. Both data architectures offer complementary perspectives; reconciling divergent findings requires future work linking individual flows to aggregate sectoral dynamics.

\paragraph{Geographic and Methodological Scope}

The analysis covers Spain's \emph{régimen común}, excluding the Basque Country and Navarre (6\% of population, 7.6\% of GDP), which operate under separate fiscal regimes. While these higher-productivity regions exhibit lower minimum-wage bite, their exclusion is unlikely to overturn the aggregate null finding for the common-regime territory. Additionally, the structural exposure measures (Youth Incidence and Monetary Gap) rely on Exponential Tilting imputation, which assumes intra-sectoral wage inequality mirrors regional inequality conditional on cell-level constraints. This assumption is plausible for broad sectors but may be less accurate for narrowly defined cells. While robustness checks using institutional proxies (Kaitz Index, Sectoral Incidence) yield qualitatively consistent results, decimal-level precision of imputed bite measures should not be over-interpreted.

\paragraph{External Validity and Policy Context}

Findings are linked to a particular institutional setting and macroeconomic environment: Spain in 2019, with a minimum wage rising from 52\% to 61\% of average wages during economic recovery, in a dual labour market with high structural youth unemployment and marked regional heterogeneity. The absence of large aggregate youth employment losses in this setting does not imply minimum wages can be raised indefinitely without consequence. We do not extrapolate to substantially higher levels (approaching 70--80\% of average wages), markedly different macroeconomic conditions (deep recessions), or countries with weaker enforcement, larger informal sectors, or lower regional mobility. Nor do we address informal employment, wage inequality, productivity, or household welfare. The methodological framework—combining census data, structural treatment definitions, and robust inference—can be fruitfully applied elsewhere, but whether similar bounds hold in other settings remains an open empirical question.


\paragraph{Further Work on Heterogeneous Effects}

Because of space constraints, this paper only reports a first-pass analysis of heterogeneous responses across regions, sectors, and outcomes. A separate companion paper is planned to examine these heterogeneous effects in greater detail, including wild-cluster bootstrap inference, alternative clustering schemes, and additional robustness checks, as well as econometric techniques specifically tailored to the analysis of treatment-effect heterogeneity.


\section{Conclusion}

This paper uses the 2019 Spanish minimum-wage reform as a quasi-experimental setting to study the employment effects of a large, discrete increase in the wage floor for young workers. Leveraging census-grade administrative tax data and structurally grounded measures of treatment intensity, we estimate the impact of a 22\% increase in the statutory minimum wage on aggregate youth employment. Across a range of specifications, including two-way fixed effects, dynamic event-study models and robust confidence intervals based on the \texttt{HonestDiD} framework, we find no evidence of \emph{substantial} net disemployment effects for young workers. Point estimates are small and often positive, and confidence intervals continue to include zero even with sizable deviations from parallel trends, consistent with minor positive or negative effects around the null.

A central message of the paper is that the estimated elasticity is highly sensitive to how minimum-wage exposure is measured. Simple institutional proxies such as the Kaitz Index or broad sectoral incidence can yield substantively different conclusions and, in some cases, suggest potentially misleading "stimulus” effects on outcomes like real sales. By contrast, structurally informed measures — Youth Incidence and the Monetary Gap — that are explicitly tied to the distribution of low wages and the implied increase in the wage bill do not point to significant adverse effects on aggregate youth employment. Our census-based perspective also helps reconcile prior micro-record evidence of higher separation risks for specific groups with the absence of job destruction in the aggregate: worker-level reallocations can occur without a net loss of formal employment capacity.

Robust inference and careful treatment of concurrent shocks are essential in this context. Strong positive pre-trends in high-bite sectors imply that conventional difference-in-differences estimators that impose exact parallel trends may misattribute part of pre-existing growth to the reform. Application of \texttt{HonestDiD} shows that the null finding is robust to a wide range of plausible departures from parallel trends and rules out large adverse employment effects under conservative assumptions. In addition, a triple-difference specification exploiting pre-existing tourism exposure indicates that a substantial portion of the 2020 employment collapse is driven by the COVID-19 shock operating through tourism-intensive sectors, rather than by the minimum-wage increase. Once this channel is accounted for, there is no evidence that the 2019 reform was a dominant driver of youth employment losses during the pandemic.

Taken together, these findings suggest that, in the institutional and macroeconomic environment prevailing in Spain in 2019 — with a minimum wage rising to around 60\% of the average wage and an economy in recovery — the labour market was able to absorb a large discrete increase in the wage floor without destroying aggregate youth employment. At the same time, our analysis is deliberately circumscribed: it does not rule out more modest disemployment effects, particularly in specific high-exposure regions, nor does it speak directly to other margins such as hours, job quality, or informality. The broader contribution of the paper is methodological. It provides a template for combining census-grade administrative data with structurally defined exposure measures, robust difference-in-differences inference under pre-trends, and explicit modelling of concurrent shocks. We hope that this blueprint will inform future evaluations of minimum-wage policies and other labour-market reforms in Spain and beyond.

\newpage
\printbibliography

\end{document}